\UseRawInputEncoding
\documentclass[runningheads]{llncs}
\usepackage[pdftex]{graphicx}
\usepackage{amsmath}
\usepackage{amssymb}
\usepackage{mathrsfs}
\usepackage{url}
\usepackage{color}
\usepackage[misc]{ifsym}

\begin{document}

\mainmatter  
\title{TSGN: Transaction Subgraph Networks for Identifying Ethereum Phishing Accounts}

\titlerunning{Transaction Subgraph Networks}

\author{Jinhuan Wang\textsuperscript{1,2}     \and
        Pengtao Chen\textsuperscript{1,2}       \and
        Shanqing Yu\textsuperscript{1,2}   \and
        Qi Xuan\textsuperscript{1,2,3}\textsuperscript{(\Letter)}
}

\institute{
\textsuperscript{1} Institute of Cyberspace Security, Zhejiang University of Technology, \\Hangzhou 310023, China \\
\textsuperscript{2} College of Information Engineering, Zhejiang University of Technology, \\Hangzhou 310023, China\\
\textsuperscript{3} PCL Research Center of Networks and Communications, Peng Cheng Laboratory, \\Shenzhen 518000, China\\
\email{xuanqi@zjut.edu.cn}
}
\authorrunning{J. Wang et al.}

\toctitle{Lecture Notes in Computer Science}
\tocauthor{Authors' Instructions}
\maketitle              
\begin{abstract}

Blockchain technology and, in particular, blockchain-based transaction offers us information that has never been seen before in the financial world. In contrast to fiat currencies, transactions through virtual currencies like Bitcoin are completely public. And these transactions of cryptocurrencies are permanently recorded on Blockchain and are available at any time. Therefore, this allows us to build transaction networks (TN) to analyze illegal phenomenons such as phishing scams in blockchain from a network perspective.
In this paper, we propose a \underline{T}ransaction \underline{S}ub\underline{G}raph \underline{N}etwork (TSGN) based classification model to identify phishing accounts in Ethereum. Firstly we extract transaction subgraphs for each address and then expand these subgraphs into corresponding TSGNs based on the different mapping mechanisms. We find that TSGNs can provide more potential information to benefit the identification of phishing accounts. Moreover, Directed-TSGNs, by introducing direction attributes, can retain the transaction flow information that captures the significant topological pattern of phishing scams. By comparing with the TSGN, Directed-TSGN indeed has much lower time complexity, benefiting the graph representation learning. Experimental results demonstrate that, combined with network representation algorithms, the TSGN model can capture more features to enhance the classification algorithm and improve phishing nodes' identification accuracy in the Ethereum networks.

\end{abstract}
\section{Introduction}

Blockchain is a distributed public ledger that is secured by blockchain technology. All transactions take place between two different public addresses and are permanently recorded on a specific blockchain built for Bitcoin. The process of securing these transactions is handled by Bitcoin miners, who use their computing power to solve complex encryption problems and validate blocks and transactions in the process~\cite{nakamoto2019bitcoin}. There is no limit to the number of Bitcoin addresses that any individual or organization can create, and there is no need to verify the identity during the process of creating an address. With the above advantages, blockchain technology has been rapidly developed and naturally introduced into the financial field. In the digital currency scenarios, the most widely used application of blockchain is cryptocurrency technology~\cite{yuan2018blockchain}, by which accounts can freely and conveniently conduct transactions with currency and information and do not have to rely on traditional third parties.

It's worth noting that the cryptocurrency market inevitably breeds many cybercrimes due to anonymity and unsupervised organization. Similarly, as the second-largest cryptocurrency platform next to Bitcoin, Ethereum has been affected by many entities/accounts engaging in illegal activities over the network, including smart Ponzi schemes, phishing, money laundering, fraud, and criminal-related activities. It is reported that phishing scams can break out periodically and are the most deceptive form of fraud~\cite{chen2020phishing}. Although the hash mechanism set up inside the blockchain can prevent transactions from being tampered with, so far, there are no available internal tools that can detect illegal accounts and suspicious transactions on the network. Thus it can be seen that cybercrimes, especially phishing scams, have become a critical issue on Ethereum and should be worthy of long-term attention and research to adopt effective countermeasures.

Generally, phishing is a social engineering attack that aims to exploit weaknesses caused by users in the system processes~\cite{khonji2013phishing}. In traditional phishing attacks, the terminal consumers will receive emails or text messages containing a malicious website whose hostname is close to the legitimate domain from a trusted entity in disguise. Once the link is clicked, phishers will use the measures provided in the link to obtain the users¡¯ sensitive information, such as usernames, passwords, and credit card details. Thus, existing researches on detecting phishing scams mainly focus on the suspected phishing website identification~\cite{feng2018application,sahingoz2019machine} and phishing text massages detection~\cite{adebowale2019intelligent,gualberto2020answer}. Compared with traditional phishing scenarios, blockchain's openness and transparency make the suspicious phishing addresses and fraudulent funds reportable and traceable. Therefore, traditional forms of phishing scams are difficult to implement on the Ethereum platform on a large scale, and the corresponding detection schemes are not suitable to migrate to the Ethereum phishing detection problem.

In order to identify phishing addresses on Ethereum, we construct transaction networks by transaction information recorded permanently on the Ethereum. Each account is accessible and their transaction history can be available freely. In the transaction networks, the nodes represent Ethereum addresses, while the edges indicate the transaction records with some attributes. Generally, each record between Ethereum accounts includes information such as transaction direction, transaction amount, and transaction timestamp.
In this paper, we propose the TSGN model to identify phishing accounts. We think of transaction direction and transaction amount as the essential attributes to build transaction networks. Based on the above, we preproccess the weighted directed transaction networks and then map these networks to subgraph network structural space. According to different pre-processing and mapping strategies, we can obtain the corresponding TSGN and Directed-TSGN for the subsequent feature extraction and detection task. Specifically, our contributions can be concluded as follow:

\begin{itemize}
\item We propose a new transaction network model, transaction subgraph networks (TSGNs). Compared with original transaction networks, our TSGN can increase the diversity of features benefiting the subsequent network algorithms.
\item We introduce different network mapping strategies to fully capture the potential structural topological information which can not be obtained easily from transaction networks.
\item We build the problem of Ethereum phishing account identification as a graph classification task. Our TSGN model can be utilized to enhance various graph classification algorithms such as manual attributes, Graph2Vec, and Diffpool.
\item We apply the new model to three transaction network datasets, and our experimental results demonstrate the effectiveness of TSGNs. The fusion of TN and TSGNs generated by different mapping strategies can increase the performance of graph classification algorithms. Directed-TSGN achieves the best performance in 7 of 9 cases. Especially, the classification result Directed-TSGN increases to 93.90\% (93.25\% for TSGN) when only Diffpool is considered, greatly improving the phishing account identification performance. More remarkably, compared with TSGN, generating Directed-TSGN needs much less time, reduced by almost one order of magnitude.
\end{itemize}

The rest of the paper is structured as follows. In section \ref{sec:related}, we make a brief description of the phishing identification and graph representation methods. In Section~\ref{sec:method}, we mainly introduce the definitions and construction methods of transaction subgraph networks. In Section~\ref{sec:experiment}, we give several feature extraction methods, which together with TSGN and Directed-TSGN are applied to three Ethereum transaction network datasets. Finally, we conclude our paper in Section~\ref{sec:conclusion}.

\section{Background and Related Work}\label{sec:related}

In this section, to supply some necessary background information, we give a brief overview of phishing detection and graph representation algorithms in graph mining.

\subsection{Phishing Identification}

Phishing scams have become a major threat to the security of Ethereum transactions. To create a good investment environment in the Ethereum ecosystem, many researchers have paid lots of attention to study the effective detection methods for phishing scams. Different from the privacy of traditional financial transaction information, the transaction records of the blockchain are freely available and contain rich attributes. Therefore, many recent studies are mainly based on transaction records. Wu et al.~\cite{wu2020phishers} proposed an approach to detect phishing scams on Ethereum by mining its transaction records. By considering the transaction amount and timestamp, this work introduced a novel network embedding algorithm called trans2vec to extract the features of the addresses for subsequent phishing identification. Chen et al.~\cite{chen2020phishing} proposed a detecting method based on Graph Convolutional Network and autoencoder to precisely distinguish phishing accounts. One can see that these methods mentioned above mainly built phishing account detection as a node classification task, which can not capture more potential global structural features for phishing accounts. Yuan et al.~\cite{yuan2020phishing} built phishing identification problem as the graph classification task, which used line graph to enhance the Graph2Vec method and achieved good performance. However, Yuan et al. only consider the structural features obtained from line graphs, ignoring the direction information, which plays a significant role in phishing scams' identification problem.
As we know, in the process of phishing fraud, the phishing funds mostly flow from multiple accounts to a specific account. From the network's perspective, the phishing nodes' local topology may be more inclined to multiple inputs and a single output. Our method takes the direction information into consideration and builds the Directed-TSGN model, revealing the topological pattern of phishing scams.

\subsection{Graph Representation}

Network, as a general modeling approach, are frequently used to study various real world systems, such as social networks~\cite{xuan2019subgraph}, traffic networks~\cite{ruan2019empirical}, protein interaction networks~\cite{borgwardt2005protein}, literature citation networks~\cite{hosseini2018analysis}, etc.. Due to its unique structure characteristics, Blockchain ecosystem is naturally modeled as transaction networks to carry out related research. Simultaneously, many graph representation methods are applied to capture the dependency relationships between objects in the Blockchain network structure. Alarab et al.~\cite{alarab2020competence} adopted Graph Convolutional Networks (GCN) intertwined with linear layers to predict illicit transactions in the Bitcoin transaction graph and this method outperforms graph convolutional methods used in the original paper of the same data. Liu et al.~\cite{liu2020graph} introduced an identify inference approach based on big graph analytics and learning, aiming to infer the identity of Blockchain addresses using the graph learning technique based on Graph Convolutional Networks. Zhang et al. \cite{zhuang2020smart} constructed a graph to represent both syntactic and semantic structures of an Ethereum smart contract function and introduced the graph neural network for smart contract vulnerability detection. According to the above works, one can find that graph representation methods can indeed be utilized to study blockchain networks and outperform in many different applications. In this work, we introduce three categories of graph representation methods such as handcrafted features \cite{xuan2019subgraph}, embedding method Graph2Vec~\cite{Narayanan2016graph2vec}, and deep learning method Diffpool~\cite{ying2018hierarchical}, to extract the features of TNs, TSGNs, and Directed-TSGNs, preparing for the subsequent phishing account identification.

\section{Methodology}\label{sec:method}

In this section, we first formulate the problem description and then present the construction detail of the transaction subgraph network model.

\subsection{Problem Description}

Generally, given a set of addresses on Ethereum, we can construct transaction network $G=(V,E,W)$, where the node set $V$ indicates the set of addresses, the edge set $E$ represents the transaction from a source address to a destination address with the transaction amounts as the weight value set $W$.

Here, we construct a set of transaction graphs for each target address $\textbf{G} = \{G_{add.1},G_{add.2},\cdots,G_{add.n}\}$, where $G_{add.i} = (V_{add.i},E_{add.i},W_{add.i},D_{add.i},y_{add.i})$ is a transaction graph of target address $i$, $V_{add.i}$ represents address $i$ and it's neighbor addresses, $E_{add.i}$ is the directed transaction set between the addresses of $V_{add.i}$ with direction set $D_{add.i}$ and weight set $W_{add.i}$, and $y_{add.i} \in Y^{|\textbf{G}|\times |\phi|}$ is the label of address $i$ and it's corresponding transaction subgraph, where $\phi$ is the label set of all target addresses. In this work. our purpose is to learn a mapping function $\mathscr{F}:\textbf{G} \rightarrow Y$ which can predict the labels of graphs in $\textbf{G}$. The label set $Y$ includes phishing addresses and normal addresses in the scenario of Ethereum phishing account identification.

\subsection{Transaction Subgraph Networks}

In this section, we introduce the detail of our transaction subgraph network model. Firstly, we give the definitions of TSGN and Directed-TSGNs as shown in the Definition \ref{def:1} and Definition \ref{def:2}, and then we elaborate the construction methods of transaction subgraph networks (TSGNs) and directed transaction subgraph networks (Directed-TSGNs), respectively.

\begin{definition}[\textbf{TSGN}]\label{def:1}
Given a transaction graph $G=(V,E,W)$, the TSGN, denoted by $T=\mathscr{L}(G)$, is a mapping from $G$ to $T=(V',E',W')$, with the node and edge sets denoted by $V'=\{t_i|i=0,1,2,\cdots\}$ and $E'\subseteq(V'\times V')$. The transaction subgraphs $t_a$ and $t_b$ will be connected if they share the common addresses or transactions in original transaction graph $G$. The $W'$ will be calculated by a weight mapping function $W' \leftarrow f(W)$.
\end{definition}

\begin{definition}[\textbf{Directed-TSGN}]\label{def:2}
Given a directed transaction graph $G$ = $(V,E,W,D)$, the Directed-TSGN, denoted by $T_D=\mathscr{F}(G)$, is a mapping from $G$ to $T_D=(V^*, E^*, W^*, D^*)$, with the node and edge sets denoted by $V^*=\{d_i|i=0,1,2,\cdots\}$ and $E^*\subseteq(V^* \times V^*)$. A directed edge will be built between two directed transaction subgraphs $d_a$ and $d_b$ when they meet the following conditions: In the original directed transaction graph $G$, (i) they share the common addresses or transactions, (ii) and form a path with the same direction. The $W^*$ will be calculated by a weight mapping function $W^* \leftarrow f'(W)$.
\end{definition}

According to the above definitions, we can see that TSGN is a variant of SGN model \cite{xuan2019subgraph} on Ethereum transaction networks. Different from SGN model, TSGN adds a network weight mapping mechanism, which can retain the transaction amount information in the original transaction network for downstream network analysis tasks. Based on TSGN model, Directed-TSGN introduces the direction information into the mapping mechanism which can capture the path of transaction behavior. Next, we will focus on demonstrating the specific construction methods.

\subsection{Constructing TSGN} \label{sec:TSGN}

\begin{figure}[h]
\centering
\includegraphics[width=\textwidth]{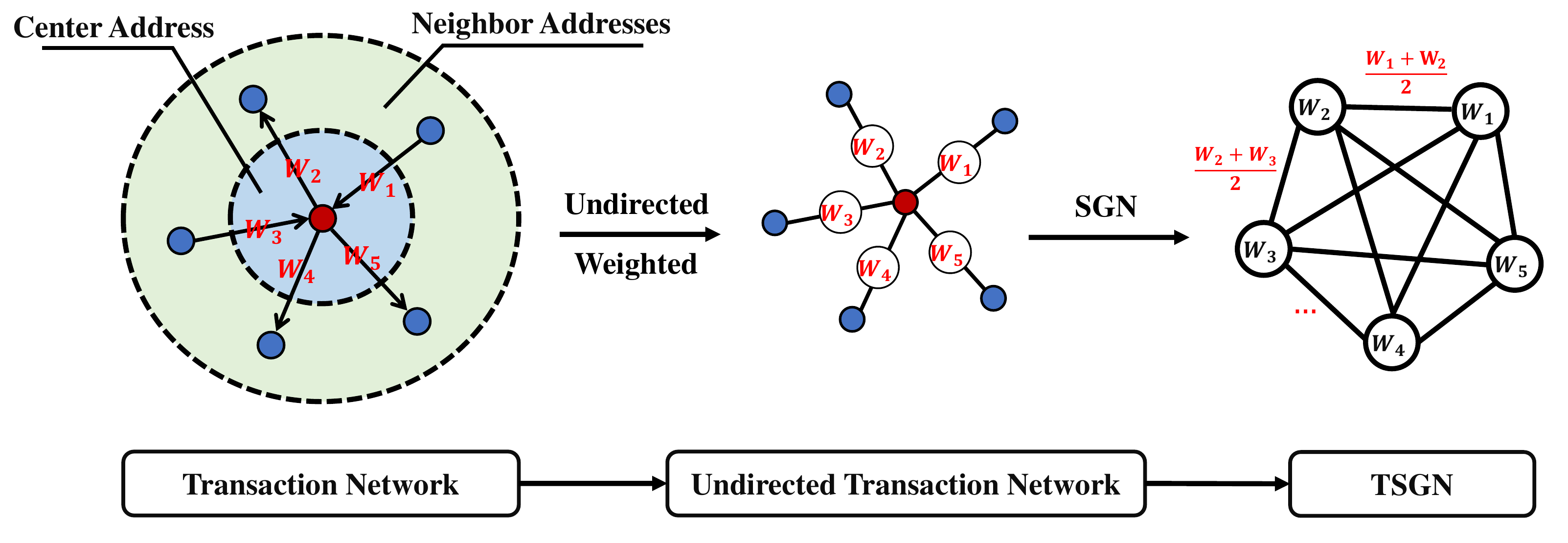}
\caption{A toy example of constructing TSGN.}
\label{TSGN}
\end{figure}


Fig.~\ref{TSGN} shows the process of constructing TSGN. Given an original transaction network composed of a center address and it's neighbor addresses, we can firstly get a plain transaction network with weight values after undirected processing. And then, we map this network into TSGN structural space. Specifically, the edges in the undirected transaction network is mapping to the nodes $W_1$, $W_2$, $W_3$, $W_4$, $W_5$ of TSGN, and then new edges are built between nodes $W_1$, $W_2$, $W_3$, $W_4$, $W_5$ because the edges of undirected transaction network share the common (red) node. We choose the mean function $f(W_{ij}) = Mean(W_i,W_j)$ as weight mapping function in Definition~\ref{def:1}, i.e., the weight of edge $(W_1,W_2)$ can be calculated as $(W_1+W_2)/2$. Of course, different weight mapping functions can be defined as required.

\subsection{Constructing Directed-TSGN} \label{sec:D-TSGN}

\begin{figure}[h]
\includegraphics[width=\textwidth]{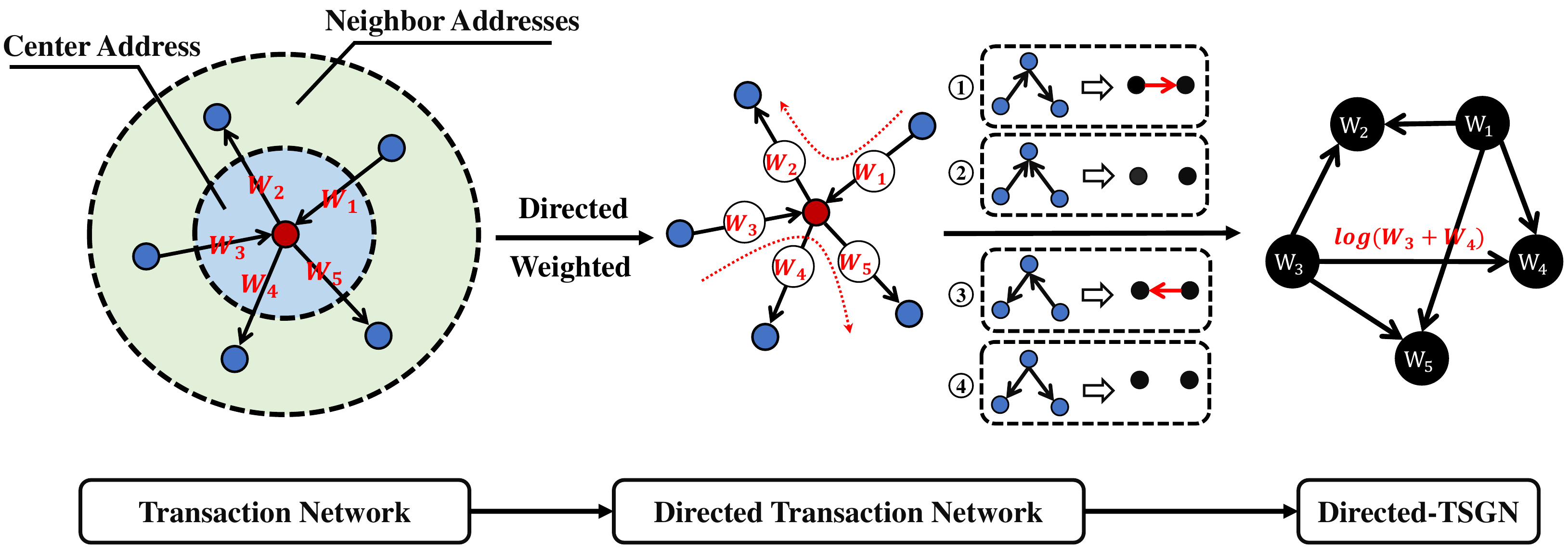}
\caption{A toy example of constructing Directed-TSGN.}
\label{D-TSGN}
\end{figure}

According to the section.~\ref{sec:TSGN}, we can find that the TSGN becomes more complex than the original transaction network, even a fully connected network, which may reduce graph mining algorithms' representation ability. Moreover, the mapping mechanism of TSGN model can not retain the direction information, which may play an important role in the following tasks.

In response to the above problems, we propose the Directed-TSGN. As shown in Fig.~\ref{D-TSGN}, the directed transaction network remains the direction and weighted attributes of the original transaction network. Similarly, the edges are mapped into the black nodes $W_1$, $W_2$, $W_3$, $W_4$, $W_5$ of the Directed-TSGN. The two red directed dashed lines indicate that the transactions $W_1$ and $W_2$ and transactions $W_3$ and $W_4$ can be seen as two continuous transaction behaviors, respectively. In other words, the edges with weights $W_1$ and $W_2$ and the edges $W_3$ and $W_4$ can form two paths with the same direction, respectively. According to the four direction mapping strategies, we can build the new edges in the Directed-TSGN. Due to the fact that \textcircled{2} and \textcircled{4} don't satisfy the requirements of constructing edges, the Directed-TSGN can limit the network size and get a relatively sparse transaction subgraph network. Here, $f'(W_{ij})=log(W_i+W_j)$ is chosen as the weight mapping function in Definition~\ref{def:2}.

\section{Experimental Evaluations}\label{sec:experiment}

\subsection{Datasets}
Ethereum, today's largest blockchain-based application, has fully open transaction data which can be easily accessed through the API of Etherscan(\url{etherscan.io}). Considering that the entire transaction network is enormous, we crawl some phishing addresses and normal addresses as the target nodes and only extract their first-order neighbor nodes from the Ethereum transaction records to construct transaction network datasets. After filtering and preprocessing the raw data, we finally got 1626 transaction networks centered on phishing nodes, and 1641 transaction networks centered on normal nodes. And then, these networks will be randomly divided so that we finally get three balanced datasets, each of which has 500 transaction networks of phishing addresses and 500 transaction networks of normal addresses. And the next experiments will be verified on these three datasets. The basic statistics of these datasets are presented in Table~\ref{tab1}.

\begin{table*}[ht]
\caption{Basic statistics of three datasets. $N_G$ is the number of graphs, \#$C_{max}$ is the number of graphs belonging to the largest class, $N_C$ is the number of classes, and \#Nodes and \#Edges are the average numbers of nodes and edges, respectively, of the graphs in the dataset.}\label{tab1}
\centering
\renewcommand{\arraystretch}{1.5}
\setlength{\tabcolsep}{4mm}{
\begin{tabular}{c|ccccc}
    \hline\hline
	Dataset &  $N_G$ & \#$C_{max}$ & $N_C$  & \#Nodes & \#Edges \\
	\hline
	EthereumG1 & 1000 & 500 & 2 & 26.003 & 25.031 \\
	EthereumG2 & 1000 & 500 & 2 & 31.650 & 30.673 \\
	EthereumG3 & 1000 & 500 & 2 & 26.338 & 25.369 \\
    \hline\hline
\end{tabular}}
\end{table*}

\subsection{Metrics}
In order to accurately evaluate the quality of each classification model, in this paper, we will use \textit{F1-Score} as a metric,

\begin{equation}
    F_1=\frac{2PR}{P+R},
\end{equation}
where \textit{P} is precision and \textit{R} is recall. \textit{F1-Score} is the harmonic mean of precision and recall, so it can more comprehensively judge the pros and cons of the classification models.

\subsection{Baselines and Experimental Setup}

For the phishing account detection problem, we transform it into a graph classification task. In order to better verify the effect of the model proposed, we adopt three typical feature extraction methods to generate graph representation, namely handcrafted attributes, Graph2Vec, and Diffpool, which are introduced in the following.

\subsubsection{Handcrafted Attributes}\label{sec:HA}
In network science, there are many classic topological attributes, which have been widely used in graph classification~\cite{li2011graph,xuan2019subgraph,wang2021sampling}, link prediction~\cite{fu2018link} and so on. In this paper, we aim to represent the networks by manually extracting the transaction network features, which are used in the downstream graph classification task. We mainly extracted 10 network features such as \emph{the number of network nodes}, \emph{the number of edges} and \emph{the average clustering coefficient}, etc.. See the Appendix for details.

\subsubsection{Graph2Vec}\label{sec:G2V}
Graph2Vec~\cite{Narayanan2016graph2vec} is the first unsupervised embedding approach for the entire networks, which is based on the embedding technique that has shown great advantages in NLP. Graph2Vec establishes the relationship between a network and the rooted subgraphs using a similar model to Doc2Vec~\cite{le2014distributed}. Graph2Vec first extracts rooted subgraphs and provides corresponding labels into the vocabulary, and then trains a skipgram model to obtain a representation of the entire network.

\subsubsection{Diffpool}\label{sec:Diffpool}
This method~\cite{ying2018hierarchical} proposed a differentiable graph pool module, which can generate hierarchical representations of graphs and can be combined with various graph neural network architectures in an end-to-end manner. Diffpool learns the distinguishable soft cluster allocation of nodes on each layer of deep GNN and maps the nodes to a set of clusters, which then form the coarse input of the next GNN layer. This method mainly solves the problem that the traditional GNN methods are flat and can't learn the hierarchical representations of graphs.

\subsubsection{Parameter Setting}\label{sec:Parameter}
The experimental part is mainly divided into two steps: the representation of the graph and the graph feature classification. In the graph feature representation part, we used the above three graph representation methods to extract features of TN, TSGN, and Directed-TSGN.
For \textit{Handcrafted Attributes}, there are no hyperparameters, just extract 10 features of each graph. For \textit{Graph2Vec}, the parameter height of the WL kernel is set to 3. Since the embedding dimension is predominant for learning performances, a commonly-used value of 1,024 is adopted. For TN and Directed-TSGN, the parameters weight and direction are set to true. But for TSGN, we set the direction to false and the weight to true. The other parameters are set to defaults: the learning rate is set to 0.025 and the epoch is set to 1000. For \textit{Diffpool}, the parameter settings of the model are the same as in \cite{ying2018hierarchical}.
We also set corresponding initialization features for different networks. For TN, the node feature is a two-dimensional vector composed of in-degree and out-degree, while for TSGN and Directed-TSGN, the node feature is a one-dimensional vector composed of the weight of the corresponding link before graph mapping. In the graph feature classifier part, each dataset is randomly split into 9 folds for training and 1 fold for testing. To exclude the random effect of fold assignment, the experiment is repeated 500 times using the random forest classifier and then the average $F_1$-$Score$ and its standard deviation recorded.

\subsection{Results}

\begin{table*}[ht]
\caption{The classification performance of different transaction subgraph network model.}\label{tab2}
\centering
\renewcommand{\arraystretch}{1.5}
\setlength{\tabcolsep}{4mm}{
\begin{tabular}{l|c|c|c}
\hline\hline
\multicolumn{1}{l|}{Datasets}& EthereumG1 & EthereumG2 & EthereumG3\\
\hline
\multicolumn{1}{l|}{ Algorithm}& \multicolumn{3}{c}{Handcrafted} \\
\hline
TN(Original)  &  $74.74\pm{3.42}$ & $76.90\pm{2.65}$ & $72.84\pm{2.92}$\\
TSGN          &  $75.25\pm{1.63}$ & $76.94\pm{2.29}$ & $73.04\pm{2.43}$\\
Directed-TSGN &  ${\bf75.35\pm{3.88}}$ & ${\bf77.50\pm{2.25}}$ & ${\bf73.95\pm{2.31}}$\\
\hline
\multicolumn{1}{l|}{ Algorithm}& \multicolumn{3}{c}{Graph2Vec} \\
\hline
TN(Original)  &  $56.45\pm{4.18}$ & $57.25\pm{1.79}$ & $61.80\pm{2.23}$\\
TSGN          &  $56.95\pm{2.42}$ & $57.85\pm{2.72}$ & $62.05\pm{3.02}$\\
Directed-TSGN &  ${\bf68.15\pm{2.26}}$ & ${\bf68.10\pm{1.28}}$ & ${\bf64.15\pm{2.48}}$\\
\hline
\multicolumn{1}{l|}{ Algorithm}& \multicolumn{3}{c}{Diffpool} \\
\hline
TN(Original)  &  $93.09\pm{1.31}$ & $89.10\pm{1.64}$ & $92.85\pm{1.09}$\\
TSGN          &  ${\bf94.35\pm{1.39}}$ & ${\bf93.64\pm{1.32}}$ & $93.25\pm{1.49}$\\
Directed-TSGN &  $93.35\pm{1.18}$ & $89.20\pm{1.53}$ & ${\bf93.90\pm{2.43}}$\\
\hline\hline
\end{tabular}}
\end{table*}

According to the above setting, we conduct some experiments on the three Ethereum datasets, and the results of phishing account identification are shown in Table~\ref{tab2}. We can find that, compared with the transaction networks (original), TSGN and Directed-TSGN models indeed has good performances in enhancing the phishing account identification. Interestingly, TSGN achieves the best classification performance 94.35\% and 93.64\%, in 2 of 9 cases based on the deep learning method Diffpool.
Overall, Directed-TSGN increases the performance of the original classification results in 7 of 9 cases. Combined with the Handcrafted Attributes method, Directed-TSGN outperforms TN, leading to an increase of 1.11\%. Directed-TSGN has an improvement over TN on all datasets, and it leads to an increase of 11.70\% when considering the Graph2Vec method.
We can see that TSGN and Directed-TSGN achieve the state-of-the-art results on the deep learning method Diffpool, which indicates that our TSGN model can further improve the representation capability of the deep learning method.

Furthermore, we record the computational times and compare the time consumption of constructing TSGN and Directed-TSGN on three datasets. The results are presented in Table~\ref{tab3}, where one can see that, the computational time of Directed-TSGN is much less than that of TSGN on each dataset, decreasing from 3 hundred seconds to less than 70 seconds. Such results suggest that, Directed-TSGN can further enhance the performance of the algorithm for phishing account identification, while also greatly improve the efficiency of the algorithms.

\begin{table*}[ht]
\caption{Time consumption (sec.) of constructing TSGNs and Directed-TSGNs.}\label{tab3}
\centering
\renewcommand{\arraystretch}{1.5}
\setlength{\tabcolsep}{4mm}{
\begin{tabular}{c|cc}
    \hline\hline
	Dataset &  TSGN & Directed-TSGN \\
	\hline
	EthereumG1 & $1.355\times{10^2}$ & 7.3687 \\
	EthereumG2 & $3.650\times{10^2}$ & 56.9006 \\
	EthereumG3 & $1.264\times{10^2}$ & 65.3633 \\
    \hline\hline
\end{tabular}}
\end{table*}

\section{Conclusion} \label{sec:conclusion}

In this paper, we present a novel transaction subgraph network (TSGN) model for phishing account identification. By introducing different mapping mechanisms into the transaction networks, we built TSGN and Directed-TSGN models to enhance the classification algorithms. Compared with the TNs, our TSGN indeed provide more potential information to benefit the phishing account identification. Considering the direction attributes, the Directed-TSGNs can retain the transaction flow information that captures the significant topological pattern of phishing scams. By comparing with the TSGN, Directed-TSGN is of a controllable scale and indeed have much lower time complexity, benefiting the network feature extraction methods to learn the network structure with higher efficiency.
Experimental results demonstrate that,  combined with network representation algorithms, the TSGN and Directed-TSGN models can capture more features to enhance the classification algorithm and improve phishing nodes identification accuracy in the Ethereum networks. In particular, when deep learning methods Diffpool is adopted to extract the features of these networks, we can achieve the state-of-the-art results on all datasets.

\section{Acknowledgments}

This work was partially supported by the National Key R\&D Program of China under Grant No. 2020YFB1006104, by the National Natural Science Foundation of China under Grant No. 61973273, and by the Zhejiang Provincial Natural Science Foundation of China under Grant No. LR19F030001.

\section{Appendix}

\begin{itemize}
\item \textbf{Number of Nodes ($N$)}: Total number of nodes in the network.
\item \textbf{Number of links ($L$)}: Total number of links in the network.
\item \textbf{Average degree ($K$)}: The mean value of links connected to a node in the network.
\item \textbf{Percentage of leaf nodes ($P$)}: A node of degree 1 is defined as a leaf node. Suppose there are totally $F$ leaf nodes in the network. Then,
\begin{equation}
P=\frac{F}{N}\,.
\end{equation}
\item \textbf{Average clustering coefficient ($C$)}: For node $v_i$, the clustering coefficient represents the probability of a connection between any two neighbors of $v_i$. Suppose that there are $k_i$ neighbors of $v_i$ and these nodes are connected by $L_i$ links. Then, the average clustering coefficient is defined as
\begin{equation}
C=\frac{1}{N}\sum_{i=1}^N\frac{2L_i}{k_i(k_i-1)}\,.
\end{equation}
\item \textbf{Largest eigenvalue of the adjacency matrix ($\lambda$)}: The adjacency matrix $A$ of the network is an $N\times{N}$ matrix, with its element $a_{ij}=1$ if nodes $v_i$ and $v_j$ are connected, and $a_{ij}=0$ otherwise. In this step, calculate all the eigenvalues of $A$ and choose the largest one.
\item \textbf{Network density ($D$)}: Given the number of nodes $N$ and the number of links $L$, network density is defined as
\begin{equation}
D=\frac{2L}{N(N-1)}\,.
\end{equation}
\item \textbf{Average betweenness centrality ($C_B$)}: Betweenness centrality is a centrality metric based on shortest paths. The average betweenness centrality of the network is defined as
\begin{equation}
C_B=\frac{1}{N}\sum_{i=1}^N\sum_{s\neq i\neq t} {\frac{n^{i}_{st}}{g_{st}}}\,,
\end{equation}
where $g_{st}$ is the number of shortest paths between $v_s$ and $v_t$, and $n^{i}_{st}$ is the number of shortest paths between $v_s$ and $v_t$ that pass through $v_i$.  	
\item \textbf{Average closeness centrality ($C_C$)}: The closeness centrality of a node in a connected network is defined as the reciprocal of the average shortest path length between this node and the others. The average closeness centrality is defined as
\begin{equation}
C_C=\frac{1}{N}\sum_{i=1}^N\frac{{n-1}}{{\sum_{j=1}^n {{d_{ij}}} }}\,,
\end{equation}
where $d_{ij}$ is the shortest path length between nodes $v_i$ and $v_j$.
\item \textbf{Average neighbor degree ($D_N$)}: Neighbor degree of a node is the average degree of all the neighbors of this node, which is defined as
\begin{equation}
D_N=\frac{1}{N}\sum_{i=1}^N \frac{1}{k_i}{\sum_{v_j\in \Omega_i} k_j}\,,
\end{equation}
where $\Omega_i$ is the neighbor set of node $v_i$, and $k_j$ is the degree of node $v_j\in{\Omega_i}$.
\end{itemize}

Note that, among the above 10 features, number of nodes ($N$), number of links ($L$), average degree ($K$) and network density ($D$) are the most basic properties of a network~\cite{xiaofan2012network}. Average clustering coefficient ($C$)~\cite{soffer2005network} is also a very popular metric to quantify the link density in ego networks. The percentage of leaf nodes ($P$) can distinguish whether a network is tree-like or rich with rings. The largest eigenvalue of the adjacency matrix ($\lambda$) is chosen since the eigenvalues  are the isomorphic invariant of a graph, which can be used to estimate many static attributes, such as connectivity, diameter, etc. Average neighbor degree ($D_N$) captures the 2-hop information. Also, centrality measures are indicators of the importance (status, prestige, standing, and the like) of a node in a network, therefore, we also use average betweenness centrality ($C_B$) and average closeness centrality ($C_C$) to describe the global structure of a network.
%
%
%
%







\end{document}